\documentclass[twocolumn]{aastex61}
\usepackage{natbib}

\newcommand{\cyanoacet}{$\mbox{HC}_3\mbox{N}$}
\newcommand{\methylcyan}{$\mbox{CH}_3\mbox{CN}$}
\newcommand{\ammonia}{$\mbox{NH}_3$}
\newcommand{\kms}{$\mbox{km~s}^{-1}$}

\submitjournal{ApJ Letters}

\shorttitle{\cyanoacet\/ maser emission in NGC\,253}
\shortauthors{Ellingsen et al.}
 \watermark{DRAFT}

\begin{document}

\title{Detection of \cyanoacet\/ maser emission in NGC\,253}

\correspondingauthor{Simon Ellingsen}
\email{Simon.Ellingsen@utas.edu.au}

\author[0000-0002-1363-5457]{Simon P. Ellingsen}
\affil{School of Physical Sciences, University of Tasmania, Hobart, Tasmania 7001, Australia}

\author{Xi Chen}
\affiliation{Shanghai Astronomical Observatory, Chinese Academic of Sciences, Shanghai 200030, China}
\affiliation{Center for Astrophysics, GuangZhou University, Guangzhou 510006, China}

\author[0000-0002-4047-0002]{Shari L. Breen}
\affiliation{Sydney Institute for Astronomy (SIfA), School of Physics, University of Sydney, NSW 2006, Australia}

\author[0000-0003-0196-4701]{Hai-hua Qiao}
\affiliation{Shanghai Astronomical Observatory, Chinese Academic of Sciences, Shanghai 200030, China}

\begin{abstract}
We report the detection of maser emission from the $J=4-3$ transition of \cyanoacet\/ at 36.4~GHz towards the nearby
starburst galaxy NGC\,253.  This is the first detection of maser emission from this transition in either a Galactic or extragalactic
source.  The \cyanoacet\/ maser emission has a brightness temperature in excess of 2500~K and is offset from the center of the galaxy by approximately 18\arcsec\/ (300~pc), but close to a previously 
reported class~I methanol maser.  Both the \cyanoacet\/ and methanol masers appear to arise near the interface between the galactic bar
and the central molecular zone, where it is thought that molecular gas is being transported inwards, producing a region of extensive low-velocity shocks.

\end{abstract}


\section{Introduction} \label{sec:intro}

NGC\,253 is the largest galaxy in the Sculptor group and is one of the closest galaxies with a nuclear starburst, at a distance of 3.4 Mpc \citep{Dalcanton+09}. The disk of the galaxy has a relatively high inclination angle to our line of sight \citep[72--78$^\circ$;][]{Puche+91} and shows strong molecular emission from a wide range of species \citep[e.g.][]{Martin+06}. It has been the object of detailed studies across the electromagnetic spectrum from radio through to $\gamma$-ray \citep[e.g.][]{Ulvestad+97,Sakamoto+11,Weiss+08,Iodice+14,Dale+09,Lehmer+13}.

NGC\,253 hosts both water masers associated with star formation within the nuclear ring \citep{Henkel+04,Hofner+06}, and OH maser emission associated with the nuclear outflow \citep{Turner+85}.  There is also evidence for the presence of an \ammonia(3,3) maser towards the centre \citep{Ott+05}.  \citet{Phillips+98a} searched for class~II 6.7-GHz methanol masers towards NGC\,253, but did not detect any emission (5-$\sigma$ sensitivity limit of $\sim$110 mJy).  Most recently, the first detection of extragalactic class~I methanol maser emission from the 36.2-GHz $4_{-1} \rightarrow 3_{0}E$ transition was reported towards NGC\,253 \citep{Ellingsen+14}.

\cyanoacet\/ is a linear molecule, first detected in interstellar space by \citet{Turner71} towards the Sgr~B2 molecular cloud.  Although it has a wide range of observable transitions at centimeter and millimeter wavelengths, and its astrophysical importance has long been recognised \citep{Morris+76}, it is relatively poorly studied.  The first interstellar detection of the $J=4-3$ transition at 36.4~GHz was towards Sgr~B2 \citep{McGee+77}.  Modelling of multi-transitional \cyanoacet\/ data towards Sgr~B2 suggests that it arises from gas with a kinetic temperature of around 20~K and number density $> 10^5$ cm$^{-3}$ \citep{Brown+85}.   More recently, observations of vibrationally excited states of \cyanoacet\/ have been used to investigate hot molecular cores (HMC).  \citet{Wyrowski+99} find the population distribution of the \cyanoacet\/ to be consistent with gas kinetic temperatures around 270~K and that it is located close to classical HMC molecular tracers such as \methylcyan.  Early observations of \cyanoacet\/ in Sgr~B2 noted the $J=1-0$ transition showed anomalously strong emission \citep{Morris+76} and it was subsequently confirmed to be a weak maser from higher angular resolution observations \citep{Hunt+99}.  However, the 9.1~GHz $J=1-0$ \cyanoacet\/ emission in Sgr~B2 is to date, the only reported interstellar maser in this molecule.  Here we report the first high resolution images of \cyanoacet\/ towards the central region of NGC\,253.

\section{Observations} \label{sec:obs}
The Australia Telescope Compact Array (ATCA) was used to observe NGC\,253 on 2014 October 10 and November 27 (project code C2879).  The array was in the 1.5A configuration (baseline lengths between 153 and 1469 m) for the October session and in the EW367 (baseline lengths between 46 and 367 m) for the November session.  The Compact Array Broadband Backend \citep{Wilson+11} was configured with 2 x 2048 MHz bands, both centred at a frequency of 36.9~GHz.  One of the 2048 MHz bands had 2048$\times$1 MHz spectral channels, while the other had two ``zoom-bands'' with spectral channels of width 31.25~kHz and a bandwidths of 128~MHz (corresponding to 4096 channels per zoom-band).  The primary purpose of these observations was to investigate the class~I methanol maser emission previously observed in NGC\,253 \citep{Ellingsen+14} and a detailed investigation of those data will be presented in a forthcoming paper.  The 2048$\times$1~MHz band includes the rest frequency of the  $J=4-3$ transition of \cyanoacet\/ (36.392332~GHz), which has 6 hyperfine components which are spread over a frequency range of 3.3~MHz.  Of the six hyperfine components only the $F=3-2$, $F=4-3$ and $F=5-4$ are expected to be significant in interstellar observations \citep{McGee+77}, and these span a frequency range of 0.13~MHz, significantly less than the 1~MHz spectral resolution of the current observations.  

The data were reduced with {\sc miriad} following the standard data reduction techniques for ATCA observations.  Amplitude calibration was with respect to Uranus and PKS\,B1921$-$293 was observed as the bandpass calibrator.  The data were corrected for atmospheric opacity and the absolute flux density calibration is estimated to be accurate to 30 per cent.  The observing strategy interleaved 10 minutes on NGC\,253 (pointing centre $\alpha = 00^{\mbox{h}}47^{\mbox{m}}33.10^{\mbox{s}}$ ; $\delta = -25^\circ 17\arcmin18.0\arcsec$ (J2000)) with 2 minute observations of a nearby phase calibrator (0116-219) before and after the target source.  The total duration onsource for NGC\,253 was around 4--5 hours for each array configuration. The data from each ATCA array configuration were calibrated and reduced independently. The strong continuum emission from the central region of NGC253 was then used to self-calibrate the data from each session, prior to continuum subtraction.  This significantly improves the signal to noise for both the continuum and line datasets.  The RMS noise levels in the \cyanoacet\/ cubes (1~MHz, or 8.2~\kms\/ channel width) was 0.4~mJy~beam$^{-1}$, 0.3 mJy beam$^{-1}$ and 0.3 mJy beam$^{-1}$ for the EW367, 1.5A and combined array data, respectively.

In order to investigate the compact emission from the \cyanoacet\/ $J=4-3$ transition, we used the Jansky Very Large Array (JVLA) in A-array configuration to make observations of this line towards NGC\,253 on 2015 August 13. Two 1-GHz bands (8-bit samplers) with frequencies centered at around 36.1 GHz and 37.1 GHz were used for these observations. Each band included 8$\times$128 MHz dual polarization sub-bands, with 128 spectral channels each of 1 MHz. The \cyanoacet\/ transition was located in a sub-band centered at 36.396 GHz. The calibration of the time-dependent antenna gains was performed by frequent observations of the quasar J0025-2602. J2253+1608 (3C454.3) was used to calibrate the bandpass response and the absolute flux density scale was determined through observations of 3C48. The pointing centre for the JVLA observations of NGC\,253 was the same as for the ATCA observations and the onsource integration time was about one hour. The visibility data were calibrated using the standard JVLA Calibration Pipeline\footnote{see https://science.nrao.edu/facilities/vla/data-processing/pipeline} with the Common Astronomy Software Applications (CASA) package. The imaging and analysis were carried out in {\sc miriad}. The synthesized beam size for the image is $0.16''\times0.07''$ with a position angle of $-28^\circ$ northwest. Similar to the ATCA data reduction, self-calibration using the continuum emission from NGC\,253 was performed prior to continuum subtraction. 

\section{Results} \label{sec:results}

\begin{deluxetable*}{cccrcccc}
\tablecaption{Summary of \cyanoacet\/ emission regions in NGC\,253} 
\tablehead{
                                   & \colhead{Right}         &                                   & \colhead{Peak}      & \colhead{Velocity} & \colhead{Integrated} &                                                     &  \\
\colhead{\cyanoacet} & \colhead{Ascension} & \colhead{Declination} & \colhead{Velocity} & \colhead{Range}   & \colhead{Flux}           & \colhead{\citeauthor{Leroy+15}} & \colhead{\citeauthor{Meier+15}} \\
\colhead{source}        & \colhead{(J2000)}     & \colhead{(J2000)}       & \colhead{(\kms)\tablenotemark{a}}   & \colhead{(\kms)\tablenotemark{a}}    & \colhead{(mJy \kms)\tablenotemark{b}}  &  \colhead{cloud\tablenotemark{c}}                          & \colhead{region\tablenotemark{c}} 
}
\startdata
A & 00:47:31.98 & -25:17:28.7 & 310 & 277--343 & 294 & 1 (1.0\arcsec) & 3 (0.7\arcsec) \\
B & 00:47:32.30 & -25:17:20.4 & 335 & 253--360 & 439 & 2 (0.8\arcsec) & 2 (1.3\arcsec)\\
C & 00:47:32.82 & -25:17:21.4 & 294 & 261--318 & 443 & 3 (0.5\arcsec) & 4 (0.5\arcsec) \\
D & 00:47:33.37 & -25:17:16.1 & 179 & 154--203 & 275 & 6 (1.1\arcsec) & 7 (0.9\arcsec) \\
E & 00:47:33.62 & -25:17:12.4 & 187 & 146--220 & 257 & 7 (0.5\arcsec) & 8 (0.8\arcsec) \\
F & 00:47:33.94 & -25:17:09.6 & 212 & 187--220 & 101 & 8 (1.7\arcsec) & 9 (1.5\arcsec) \\
G & 00:47:34.13 & -25:17:12.7 & 212 & 187--261 & 166 & 9 (1.0\arcsec) & 10 (0.5\arcsec) \\
\enddata
\tablenotetext{a}{Note that all the velocities in this paper are given in the Barycentric reference frame, which for NGC\,253 are 7.24~\kms\/ lower than the equivalent velocity in the local-standard of rest (LSR) frame.}
\tablenotetext{b}{ The integrated flux density and velocity range reported are measured over the synthesised beam of the ATCA combined array data at the location given (Figure~\ref{fig:hc3n}).}
\tablenotetext{c}{ The value in brackets gives the offset from the measured \cyanoacet\/ position.}
\label{tab:hc3n}
\end{deluxetable*}

\begin{figure}
 \begin{center}
    \begin{minipage}[t]{1.0\textwidth}
      \includegraphics[angle=270,scale=0.40]{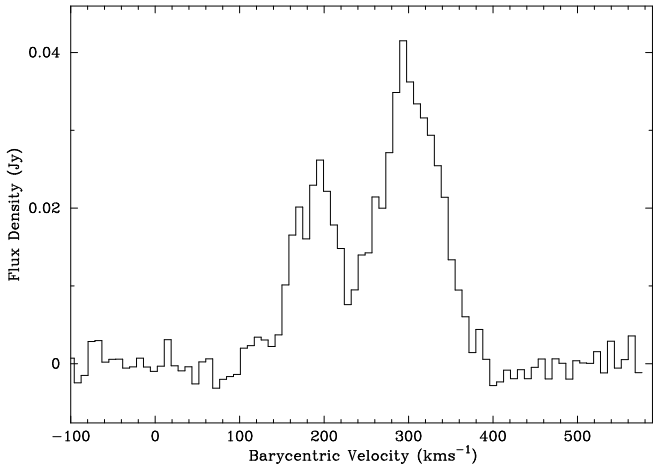}
    \end{minipage}
 \end{center}
\caption{ATCA spectrum of the integrated 36.4~GHz \cyanoacet\/ emission from NGC\,253, extracted from the combined array image cube.} \label{fig:spec}
\end{figure}

\begin{figure}
\epsscale{1.2}
\plotone{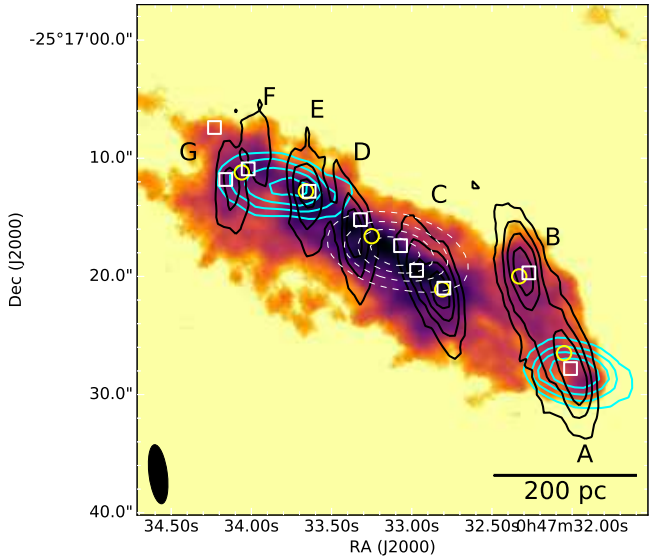}
\caption{ATCA combined EW367 and 1.5A array observations of the \cyanoacet\/ emission towards NGC\,253 (black contours at 20, 40, 60 and 80\% of 497 mJy \kms\/ beam$^{-1}$, synthesised beam 5.0\arcsec $\times$ \ 1.5\arcsec, image RMS 15 mJy \kms\/ beam$^{-1}$). 36~GHz continuum emission (dashed white contours at 20, 40, 60 and 80\% of 95 mJy beam$^{-1}$) and 36.2~GHz methanol emission (cyan contours at 20, 40, 60 and 80\% of 310 mJy \kms\/ beam$^{-1}$) towards the central molecular zone in NGC\,253 \citep{Ellingsen+14}.  The background image is the integrated CO ($J=2-1$) emission from \citet{Sakamoto+11} shown on a logarithmic scale.  The white squares mark the location of molecular clouds identified by \citet{Leroy+15} and the yellow circles the location of \ammonia\/ sources \citep{Lebron+11}.} \label{fig:hc3n}
\end{figure}

The continuum subtracted data from the two ATCA array configurations were combined during imaging, to improve both the overall sensitivity and the range of angular scales for which information is available.  Figure~\ref{fig:spec} shows the integrated spectrum of the integrated 36.4~GHz \cyanoacet\/ emission, extract from the combined array ATCA image cube. The integrated emission is 4.6 Jy~\kms\/ and covers a Barycentric velocity range from approximately 150 -- 380~\kms.  Figure~\ref{fig:hc3n} shows the spatial distribution of the 36.4~GHz \cyanoacet\/ emission from the ATCA observations, along with a range of other molecular tracers.  Recent ALMA observations of NGC\,253 have given us high resolution (2.0\arcsec $\times$ 1.5\arcsec) information on the molecular gas content in the central regions of NGC\,253 in a wide range of different molecules and transitions \citep{Leroy+15,Meier+15}.  \citeauthor{Leroy+15} identified 10 regions with significant volumes of dense gas and these are marked with white squares in Figure~\ref{fig:hc3n}.  The highest angular resolution ATCA \cyanoacet\/ observations have an angular resolution of around 2\arcsec, which corresponds to a linear scale of approximately 30~pc (assuming a distance of 3.4~Mpc to NGC\,253) and the combined array ATCA observations show 7 distinct regions of emission.  We have labeled these regions A through G in order of increasing right ascension (Figure~\ref{fig:hc3n}).  Table~\ref{tab:hc3n} summarises the properties of the 7 \cyanoacet\/ emission regions and their relationship to recent ALMA molecular line observations \citep{Leroy+15,Meier+15}. The \cyanoacet\/ emission regions are generally offset by 0.5 -- 1\arcsec\/ from the ALMA-identified molecular cloud complexes, this is comparable to the relative astrometric uncertainty between the two datasets and so they are either associated, or in close proximity. The 7 \cyanoacet\/ regions encompass 9 of the 10 molecular complexes identified in the ALMA observations of \citep{Leroy+15}.  Figure~\ref{fig:hc3n} shows that \cyanoacet\/ region A is is close to the location of the strongest 36.1~GHz class~I methanol maser in NGC\,253 \citep{Ellingsen+14}, with an offset between the two of 0.75\arcsec, corresponding to a linear separation of approximately 12~pc.

\begin{figure*}
\epsscale{1.1}
\plotone{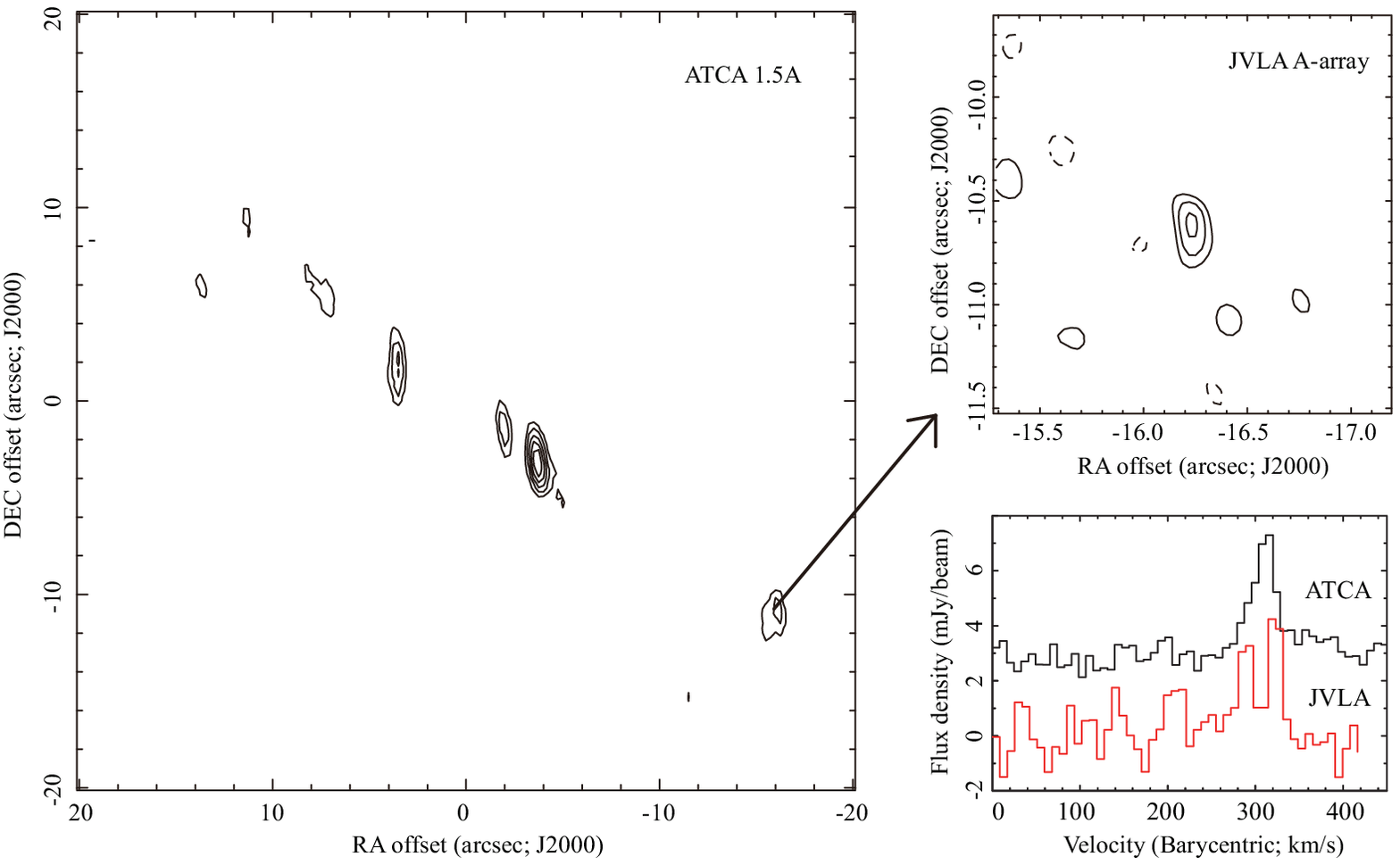}
\caption{ATCA 1.5A array and JVLA A-array observations of the integrated $J=4-3$ \cyanoacet\/ emission.  The contour levels in the ATCA image are at -3,3,5,7,9 and 11 times the RMS in the image (15 mJy beam$^{-1}$ \kms) The contour levels in the JVLA image are at -3, 3, 4 and 5 times the RMS in the image (25 mJy beam$^{-1}$ \kms).  The JVLA observations covered the same field of field of view as the ATCA, but emission was only detected towards one of the emission sites observed at lower angular resolution.} \label{fig:vla}
\end{figure*}

In addition to the images formed using the combined data from the two ATCA configurations we also examined the images of the \cyanoacet\/ emission from each configuration separately.  The 1.5A array observations have a synthesised beam size of 2.6\arcsec $\times$ 0.6\arcsec\/ and at this resolution, much of the \cyanoacet\/ emission is resolved (Figure~\ref{fig:vla}).  The two innermost \cyanoacet\/ regions which are strongest in the combined image remain the strongest in the higher resolution image, however, what is more striking is that some emission from the weaker, outer sites remains and this emission correlates closely with the 36.1~GHz class I methanol maser emission.  The JVLA A-array observations have an order of magnitude higher angular resolution (0.1\arcsec) and in this data the \cyanoacet\/ detection is detected from only one location.  The detection is statistically marginal, at 5$\sigma$ in the integrate intensity (moment 0) image and 4$\sigma$ in the strongest spectral channel, however, we are confident that the detection is robust, as it matches both the position and velocity of the peak observed in the ATCA observations (Figure~\ref{fig:vla}).  For the assumed distance to NGC\,253 an angular resolution of 0.1\arcsec\/ corresponds to a linear scale of 1.6~pc.

\section{Discussion}

A useful place to commence our investigation of the detected \cyanoacet\/ emission in NGC\,253 is to compare our results with those from Galactic sources. However, there are relatively few published observations of \cyanoacet\/ in Galactic sources and even fewer which report data on the $J=4-3$ transition. One of the best studied sources is the star forming complex Sgr\,B2, which has been observed in a wide range of \cyanoacet\/ transitions \citep[e.g.][]{McGee+77,Hunt+99}.  Early observations of this source noted that the $J=1-0$ transition showed an anomalously high intensity compared to the higher $J$ lines and weak maser emission in this line was subsequently confirmed by high resolution observations \citep{Hunt+99}.  However, to our knowledge, the $J=1-0$ emission in Sgr\,B2 is to date the only \cyanoacet\/ maser emission reported in the published literature.

The $J=4-3$ emission in Sgr B2 has a peak intensity of 1.25 K and a full-width to half maximum of 25~\kms\/ \citep{McGee+77}, which corresponds to an integrated flux density of approximately 660 Jy~\kms\/ (assuming 20~Jy/K for the Parkes 36 GHz observations).  Assuming a distance of 7.9~kpc for Sgr\,B2 \citep{Reid+09c} and 3.4~Mpc for NGC\,253 \citep{Dalcanton+09}, we find that the total \cyanoacet\/ $J=4-3$ emission in NGC\,253 (integrated flux density 4.6~Jy~\kms) is approximately 1300 times more luminous than that observed from the Sgr\,B2 complex.  \citet{Wyrowski+99} present data from the vibrational ground state of the $J=4-3$ \cyanoacet\/ transition towards the HMC source G\,10.47+0.03, however, the focus of their paper is the vibrationally excited states, so few details are given of this transition, however, it appears that the emission is more than an order of magnitude weaker than that observed towards Sgr\,B2 in the same transition.  The star formation rate in NGC\,253 \citep[4.2 M$_\sun$~yr$^{-1}$;][]{Sanders+03}, is significantly higher than in the Milky Way \citep[0.7--1.5 M$_\sun$~yr$^{-1}$;][]{Robitaille+10}, and approximately 50\% of this is concentrated in the central region of NGC\,253. However, even though Sgr\,B2 is sometimes described as a mini-starburst, it is clear that the luminosity of the \cyanoacet\/ in NGC\,253 is much larger than can be explained through linear scaling with the star formation rate.

The ALMA observations of \citet{Meier+15} included the $J=11-10$ \cyanoacet\/ transition at 100.1 GHz and it shows a similar distribution to that seen in the ATCA combined array image (Figure~\ref{fig:hc3n}).  The spatial distribution of the \cyanoacet\/ is also similar to that observed in HCN and some other thermal molecular lines.  Comparing the \cyanoacet\/ emission in Figures~\ref{fig:hc3n} and \ref{fig:vla}, it is clear that the much of the emission is resolved in the higher angular resolution observations, suggesting it is thermal.  However, some of the \cyanoacet\/ emission associated with region A is much more compact.  Considering just the \cyanoacet\/ $J=4-3$ emission detected in the ATCA observations of region A, at an angular resolution of 1.2\arcsec\/ (1.5A array observations) it has an integrated flux density of 285 mJy~\kms, corresponding to a luminosity a factor of 80 greater than Sgr\,B2.  At the angular resolution of the JVLA A-array observations we have integrated emission of approximately 200~mJy~\kms\/ from a region of 0.1\arcsec, corresponding to a luminosity more than 55 times that of Sgr\,B2, but arising from a smaller region (0.1\arcsec\/ corresponds to a linear scale of 1.6~pc for a distance of 3.4~Mpc).  From the JVLA A array observations we can put a lower limit on the brightness temperature of the \cyanoacet\/ emission of $>$ 2500~K integrated over a 50~\kms\/ velocity range ($>$ 560~K in the peak spectral channel).  \citet{Ott+05} found kinetic temperatures in the range 150--200 K for the molecular clouds in the central region of NGC\,253 from ATCA \ammonia\/ observations (angular resolution 19\arcsec $\times$ 5\arcsec), consistent with H$_2$ excitation temperatures of $\sim$200 K for the NGC\,253 molecular gas determined from infrared observations \citep{Rigopoulou+02}.  In contrast, based on data from a large number of thermal molecular transitions \citet{Meier+15} estimate kinetic temperatures around 80~K in the inner molecular clouds in NGC\,253 (those associated with \cyanoacet\/ regions C and D). They also measured excitation temperatures in the range 10--75 K for the NGC\,253 central molecular clouds from two SO transitions in their ALMA data, and radiative transfer modelling inferred similar results.  For \cyanoacet\/ the critical density is $5 \times 10^{5}$ cm$^{-3}$ \citep{Green+78}, approximately an order of magnitude higher than the density inferred through modelling of line intensity ratios in NGC\,253 by \citet{Meier+15}.  Transitions with critical densities higher than the gas density will have excitation temperatures lower than the kinetic temperature, so we expect the excitation temperatures for \cyanoacet\/ in NGC\,253 to be $<$ 200~K.  \citet{Meier+15}  also found that the excitation temperatures in molecular gas at larger galactocentric radii were lower than for those closer to the nucleus.  Hence, the brightness temperatures we measured for the compact \cyanoacet\/ $J=4-3$ emission in region A are significantly in excess of the thermal or excitation temperatures expected in NGC\,253.  This is strong evidence that some of the $J=4-3$ emission from \cyanoacet\/ region A in NGC\,253 is being produced by a non-thermal process (i.e. it is a maser).

The most widely studied extragalactic masers are the 22~GHz water megamasers and the 1.6~GHz OH megamasers.  The water megamasers are very compact and have been used to investigate kinematics and dynamics of the circumnuclear region of active galaxies at very high angular resolution \citep[e.g.][]{Argon+07}.  The OH megamasers are less well studied at high angular resolution, but it is clear that some of the maser emission arises in larger scale (10-100~pc) diffuse regions \citep[e.g.][]{Rovilos+03}.  This also appears to be the case for extragalactic H$_2$CO masers \citep{Baan+17} and class~I methanol \citep{Chen+15}.   The majority of the $J=4-3$ \cyanoacet\/ emission in NGC\,253 is likely thermal, but it may be that some of the emission which is resolved in the higher angular data presented here is from larger scale, diffuse maser regions, with the emission detected in the JVLA A-array observations representing just the most compact components of a larger maser emission region.

\citet{Brown+85} modeled the \cyanoacet\/ emission in Sgr\,B2 based on multi-transition single dish data.  They found that at modest densities ($n_{H_2} \sim 10^{2.5}$ -- $10^{5}$ cm$^{-3}$), the $J=1-0$ transition can exhibit maser emission for a wide range of temperatures.  Inversion in higher $J$ transitions was shown to occur for a more restricted range of temperatures and densities, shifting to higher values of each as $J$ increased.  This is because \cyanoacet\/ is inverted under conditions where neither collisional nor radiative processes dominate and more transitions invert as the temperature increases (radiative input) and the density increases (collisional input).  The highest energy transition for which they found inversion in their model was the $J=4-3$ for which they predict maser emission only for kinetic temperatures $>$ 50~K and densities in a narrow range around $n_{H_2} \sim 10^{4.2}$ cm$^{-3}$.  These temperatures and densities are broadly consistent with the values measured in the molecular clouds in NGC\,253 by \citet[][80~K, 10$^{4.75}$~cm$^{-3}$]{Meier+15} and others.

The \cyanoacet\/ maser emission in NGC\,253 arises within molecular cloud complex 1 from the \citet{Leroy+15} study.  This region covers a larger spatial scale and shows broader line profiles in both quiescent molecular gas and dense gas tracers than the other molecular cloud complexes they identify (see their figure~10).  The NGC\,253 molecular clouds are highly supersonic and have surface densities which are comparable to those observed in the Milky Way Galactic center clouds, but are observed out to much larger galactocentric radii \citep{Leroy+15}, and it is speculated that these characteristics result in more efficient star formation. The \citet{Meier+15} study examined the ratio of HNCO to SiO towards a number of regions in NGC\,253.  SiO is thought to be produced from sputtering of grains in high velocity shocks, while HNCO is produced by dust ice mantle sublimation which requires weak shocks.  \citet{Meier+15} found the HNCO:SiO ratio to be enhanced in the molecular clouds at higher galactocentric radii (their regions 2, 3, 9 and 10 which correspond to \cyanoacet\/ regions B, A, F and G respectively).  They hypothesize that these are regions where weak shocks dominate and are similar to conditions observed in bar shock regions of other nearby spirals such as IC342 and Maffei 2, suggesting that they may be high-priority targets for future searches.

\section{Conclusions}

High resolution observations of the 36.4~GHz $J=4-3$ transitions of \cyanoacet\/ in NGC\,253 have detected weak maser emission offset from the nucleus of the galaxy.  This emission appears to be associated with a region of the galaxy close to the inner edge of the bar, where there is a significant abundance of molecular gas and widespread low-velocity shocks.  The present observations of $J=4-3$ \cyanoacet\/ maser emission are limited to a single source, so we are limited in the inferences we can draw. However, based on the information in the literature it would appear that lower $J$ transitions of \cyanoacet\/ may show more widespread maser emission in NGC\,253 and that sources where bar shocks are observed may be likely targets to find other \cyanoacet\/ masers.  Understanding the physical significance of \cyanoacet\/ masers requires observations of additional sources and transitions to enable a better characterisation of the emission properties and their relationship with the host galaxy and we encourage further studies to facilitate this.

\acknowledgments

The Australia Telescope Compact Array is part of the Australia Telescope National Facility which is funded by the Commonwealth of Australia for operation as a National Facility managed by CSIRO. X.C. acknowledges support from National Natural Science Foundation of China (NSFC) through grant NSFC 11590781. This research has made use of NASA's Astrophysics Data System Abstract Service.  This research has made use of the NASA/IPAC Extragalactic Database (NED) which is operated by the Jet Propulsion Laboratory, California Institute of Technology, under contract with the National Aeronautics and Space Administration.

\end{document}